\documentclass[conference]{IEEEtran}
\IEEEoverridecommandlockouts
\usepackage[T1]{fontenc}
\usepackage{cite}
\usepackage{amsmath,amssymb,amsfonts}
\usepackage{algorithmic}
\usepackage{graphicx}
\usepackage{textcomp}
\usepackage{xcolor}
\usepackage{url}
\def\BibTeX{{\rm B\kern-.05em{\sc i\kern-.025em b}\kern-.08em
    T\kern-.1667em\lower.7ex\hbox{E}\kern-.125emX}}
\usepackage{booktabs,array,tikz}
\usetikzlibrary{positioning}
\usepackage[hidelinks]{hyperref}
\providecommand{\tightlist}{\setlength{\itemsep}{0pt}\setlength{\parskip}{0pt}}
\begin{document}

\IEEEoverridecommandlockouts

\IEEEpubid{\makebox[\columnwidth]{Accepted to special session track.\hfill} \hspace{\columnsep}\makebox[\columnwidth]{ }}

\title{
{A Combined Streaming FPGA Architecture for Client-Side CKKS Encoding and Encryption}
}

\author{\IEEEauthorblockN{{Joseph Kurina}}
\IEEEauthorblockA{
\textit{{University of Washington}}\\
{Bothell, Washington, USA} \\
{kurinj@uw.edu}}
\and
\IEEEauthorblockN{{Bogdan Pasca}}
\IEEEauthorblockA{
\textit{{Altera Corporation}}\\
{Paris, France} \\
{bogdan.pasca@altera.com}}
\and
\IEEEauthorblockN{{Suleyman Demirsoy}}
\IEEEauthorblockA{
\textit{{Altera Corporation}}\\
{London, U.K.} \\
{suleyman.demirsoy@altera.com}}
\and
\IEEEauthorblockN{{Sunwoong Kim}}
\IEEEauthorblockA{{\textit{Rochester Institute of Technology}}\\
{Rochester, New York, USA} \\
{sskeme@rit.edu}}
}

\maketitle
\begin{abstract}
Homomorphic encryption allows a computing service to process encrypted data, but the client must first encode and encrypt its inputs. Sharing hardware between floating-point encoding and modular encryption reduces area but requires those transform roles to execute at different times. We present a streaming FPGA architecture for CKKS encoding and encryption that computes the shared floating-point encoding transform once and performs the integer transforms for six residue moduli in parallel. Dedicated transform pipelines exchange intermediate results on chip rather than through off-chip memory. DSP Builder generates the transform RTL, and Altera HLS IP Gen compiler compiles the surrounding SYCL pipeline into IP integrated with DMA and control logic in an accelerator functional unit. At polynomial degree \(N=8192\), the Altera Agilex 7 implementation delivers 5,143 complete ciphertexts per second at 471 MHz. One SEAL-Embedded CPU worker on the same workstation delivers 113 ciphertexts per second for real-valued inputs using all 4,096 slots. Measurements cover prepared-input computation through complete ciphertext construction, including FPGA submission and output delivery.
\end{abstract}


\section{Introduction}

Homomorphic encryption (HE) allows a computing service to operate on encrypted data without decrypting it \cite{cheon2021introduction}. Applications include privacy-preserving neural-network inference \cite{gilad2016cryptonets}, image processing \cite{kim2024accelerating}, and speech recognition \cite{elworth2022hekws}. The data owner retains the secret key, prepares encrypted inputs, and decrypts the returned results. The Cheon, Kim, Kim, and Song (CKKS) scheme supports approximate arithmetic on packed real or complex values \cite{draft1}. Packing places many values in one ciphertext so that the service can process them together.

Preparing that ciphertext requires substantial computation before the service can begin. The client maps a message vector to polynomial coefficients using a complex-valued transform, scales and rounds them to integers, and encrypts the polynomial. A residue number system (RNS) represents the large ciphertext modulus through several smaller primes, requiring modular arithmetic for every prime \cite{draft2}. Client-side encoding and encryption therefore combine floating-point and integer workloads, whereas most existing HE hardware research addresses server-side ciphertext operations \cite{draft9, draft17, draft18, draft19, draft20, zhu2025dahe, kim2019fpga, kim2020hardware}.

Existing client accelerators balance arithmetic reuse against concurrent execution. Aloha-HE \cite{draft4} reuses butterfly units to reduce area, while ABC-FHE \cite{draft15} uses parallel streaming cores with shared Fourier and modular transform arithmetic. The accelerator proposed in this paper uses a workstation-attached FPGA with dedicated parallel number theoretic transforms (NTTs) and shares only the modulus-independent inverse fast Fourier transform (IFFT). This partition removes competition for transform engines between moduli and between the secret-key and plaintext-plus-error paths. The resulting architecture trades arithmetic reuse for concurrent encoding and encryption. Encoded coefficients pass into encryption through on-chip buffers, avoiding an intermediate off-chip memory transfer. Here, \emph{client-side} identifies the data-owner operation performed by the workstation, which supplies parameters, keys, and randomness, and assembles the ciphertext.

The integration challenge is to preserve coefficient order across transforms and prevent input or output storage from being overwritten while still in use. SYCL expresses device computation in C++, with FPGA extensions for typed on-chip FIFOs called pipes and imported register-transfer-level (RTL) components \cite{draft7}. We use these mechanisms to connect RTL developed with Altera's DSP Builder \cite{DSPBuilder} to SYCL polynomial-arithmetic tasks. Altera HLS IP Gen compiler \cite{HLSIPGen} compiles the tasks and their pipes into one hardware pipeline. Import buffers store transform outputs, pipes carry ordered coefficients, and the host collects complete ciphertexts after accelerator completion. Each transform transfers four coefficients together through four parallel data paths, called \emph{coefficient lanes}. These lanes carry parts of one polynomial. They do not encrypt four messages independently.

The main contributions are:
\begin{enumerate}
\def\labelenumi{\arabic{enumi}.}
\tightlist
\item A streaming symmetric CKKS architecture at \(N=8192\) with one IFFT operating on complex doubles through four coefficient lanes, and six physical 30-bit RNS pipelines, each with two dedicated NTT engines.
\item A hybrid SYCL/RTL implementation generated with Altera HLS IP Gen compiler, with on-chip transform buffering and a host-memory DMA interface inside an accelerator functional unit (AFU).
\item An evaluation against one native SEAL-Embedded \cite{draft3} worker running on the same host CPU.
\end{enumerate}

The accelerator project repository is \url{https://github.com/joekurina/FPT26_Agilex7_CKKS_Accelerator}.

\section{Background and Related Work}

\subsection{CKKS Encoding and Symmetric Encryption}

CKKS represents plaintexts and ciphertext components in the negacyclic ring \(R_Q=\mathbb{Z}_Q[X]/(X^N+1)\), where \(N\) is a power of two and \(Q=\prod_i q_i\) is a product of pairwise-coprime primes. Polynomial coefficients are integers modulo \(Q\). Multiplication wraps powers of \(X^N\) to \(-1\). RNS replaces arithmetic modulo the large \(Q\) with independent computations modulo each \(q_i\) \cite{draft2}. We call each polynomial's representation modulo one prime an RNS \emph{limb}.

Encoding takes \(N/2\) complex plaintext values, called slots, and expands them into a conjugate-symmetric vector of length \(N\). A negacyclic IFFT maps this vector to polynomial coefficients. Scaling by \(\Delta\) and rounding produces an encoded integer plaintext polynomial \(m\), which is reduced modulo each prime \cite{draft1,draft2}. The scale controls the representation of fractional values.

Symmetric encryption combines \(m\) with a small error polynomial \(e\), the secret-key polynomial \(s\), and a uniform polynomial \(a\) sampled in NTT representation. The NTT is the modular counterpart of the Fourier transform. It enables polynomial multiplication to be performed by efficient point-wise multiplication of coefficients. With \(i\) indexing the RNS modulus, define \(\widehat{m+e}_i=\operatorname{NTT}_{q_i}(m+e)\) and \(\widehat{s}_i=\operatorname{NTT}_{q_i}(s)\). The ciphertext components in NTT representation are
\begin{align}
 c_{1,i} &= a_i
 \label{eq:c_1}, \\
 c_{0,i} &= \widehat{m+e}_i-\widehat{s}_i\odot a_i \pmod{q_i},
 \label{eq:c_0}
\end{align}
where \(\odot\) denotes point-wise coefficient multiplication. All per-modulus pairs \((c_{0,i},c_{1,i})\) together represent one ciphertext \((c_0,c_1)\in R_Q^2\). The two NTTs in \eqref{eq:c_0} can run independently, and each modulus admits a separate computation.

Both transforms use \(O(N\log N)\) butterfly arithmetic, where each butterfly combines a small group of coefficients. The IFFT operates on complex doubles, whose real and imaginary components are each stored as a 64-bit double-precision floating-point value. The NTT operates in \(\mathbb{Z}_{q_i}\) and uses a primitive \(2N\)-th root of unity for the negacyclic transform \cite{draft6}. At \(N=8192\), each radix-2 transform has 13 stages. Our configuration uses six 30-bit primes satisfying \(q_i\equiv1\pmod{2N}\). Modular multiplication outside the imported NTT uses Barrett reduction \cite{draft5}.

\subsection{Related Work}

Microsoft SEAL implements CKKS in software \cite{draft8}. SEAL-Embedded, the baseline used here, reduces the memory requirements of client-side CKKS operations for embedded processors \cite{draft3}. Hardware clients include CHOCO-TACO for encrypted compute offloading \cite{draft12}, RACE and RISE for RISC-V systems \cite{draft13,draft14}, and the NTT accelerator integrated with SEAL-Embedded by Di Matteo \emph{et al.} \cite{draft26}. Software and protocol approaches also reduce client work by exploiting encoding redundancy \cite{draft25} or using a bootstrapping-assisted client protocol \cite{draft16}.

Client accelerators make different choices about supported operations and arithmetic reuse. Lee \emph{et al.} provide configurable CKKS encryption and decryption architectures \cite{draft11}. Aloha-HE shares one double-precision FFT butterfly and three modular NTT butterflies across encoding, encryption, decryption, and decoding to reduce area \cite{draft4}. ABC-FHE uses two reconfigurable streaming cores that share arithmetic between FFT and NTT modes \cite{draft15}. Section~\ref{sec:prior-comparison} compares these designs with our dedicated-transform partition and distinguishes their measurement boundaries.

HEXL-FPGA implements HE arithmetic primitives in SYCL, including NTT and key switching, rather than focusing on client-side encoding and encryption \cite{IntelHEXLFPGA}. Our design combines DSP Builder-generated RTL transforms with SYCL kernels for these client operations.

\section{Accelerator Architecture}

The accelerator shares one floating-point IFFT across six dedicated modular encryption pipelines.
Table~\ref{tab:design-parameters} summarizes the fixed configuration. 
\begin{table}[t]
\centering\footnotesize
\caption{Accelerator Parameters and Measured Clock}
\label{tab:design-parameters}
\setlength{\tabcolsep}{2pt}
\begin{tabular}
{>{\raggedright\arraybackslash}p{0.56\columnwidth}>{\raggedright\arraybackslash}p{0.37\columnwidth}}
\toprule
Parameter & Value \\
\midrule
Polynomial degree \(N\) & 8,192 \\
Complex slots & up to 4,096 \\
Transform stages & 13 \\
Coefficients per transfer & 4 \\
Transfers per polynomial & 2,048 \\
RNS moduli / physical modulus pipelines & 6 / 6 \\
Modulus width & 30 bits \\
IFFT arithmetic & Complex double precision \\
Pipe residues & 32 bits \\
Shared converted coefficients & Signed 64 bits \\
Concurrent complete operations & 1 \\
NTT engines & 12 \\
Measured application clock & 471 MHz \\
\bottomrule
\end{tabular}
\end{table}%

\subsection{Overview and Design Parameters}

Fig.~\ref{fig:full-accelerator-overview} shows the complete dataflow and expands one of the modulus pipelines.
\begin{figure*}[t]
\centering
\begin{tikzpicture}[x=1cm,y=1cm,>=stealth,
  box/.style={draw,align=center,font=\footnotesize,minimum height=0.85cm,inner sep=3pt},
  mem/.style={box,fill=black!7},
  finite/.style={box,fill=white},
  rtl/.style={box,very thick,fill=black!7},
  flow/.style={->,thick},
  labeltext/.style={font=\footnotesize,align=center,fill=white,inner sep=1.5pt}]
\node[anchor=west,font=\small\bfseries] at (0,0.95) {(a) Complete host--FPGA dataflow};
\draw[dashed,black!60] (3.4,0.5) rectangle (16.65,-4.2);
\node[anchor=east,fill=white,font=\footnotesize] at (16.45,0.5)
  {AFU: DMA/control and the SYCL pipeline compiled by Altera HLS IP Gen 2026.1};
\node[mem,text width=1.6cm,minimum height=1.4cm] (hostin) at (1.0,-0.73)
  {Host input\\Vector, $e$, $s$, $c_1$\\Prepared in\\host memory};
\node[mem,text width=0.9cm,minimum height=4.15cm] (platform) at (2.75,-2.05)
  {OFS\\FIM\\+\\PIM};
\node[finite,text width=1.1cm] (inmem) at (4.2,-0.73) {DMA\\reader};
\node[finite,text width=1.35cm,minimum height=1.1cm] (entry) at (6.2,-0.73)
  {Parser\\Input FIFO\\Distributor};
\draw (7.15,0.18)--(12.75,0.18)--(12.75,-2.55)--(9.7,-2.55)--(9.7,-1.47)--(7.15,-1.47)--cycle;
\node[anchor=west,labeltext] at (7.3,0.18) {Shared IFFT task};
\node[rtl,text width=1.5cm,minimum height=1.1cm] (ifft) at (8.25,-0.73)
  {Imported IFFT\\4 coefficient\\lanes};
\node[finite,text width=2.65cm,minimum height=1.1cm] (convert) at (11.05,-0.73)
  {Scale $\Delta/N$ $\to$ round\\$\to$ signed 64-bit\\$\to$ add error $e$};
\node[mem,text width=2.5cm,minimum height=1.05cm] (signedfifo) at (14.75,-0.73)
  {Signed FIFO\\4 $\times$ 64-bit\\coefficients / entry};
\node[finite,text width=2.5cm] (fanout) at (14.75,-2.05)
  {Broadcast\\to 6 moduli};
\draw[flow] (hostin.east)--(platform.west |- hostin.east);
\draw[flow] (platform.east |- inmem.west)--(inmem.west);
\draw[flow] (inmem.east)--(entry.west);
\draw[flow] (entry.east)--(ifft.west);
\draw[flow] (ifft.east)--(convert.west);
\draw[flow] (convert.east)--(signedfifo.west);
\draw[flow] (signedfifo.south)--(fanout.north);
\node[mem,text width=1.7cm,minimum height=0.7cm] (errorfifo) at (8.6,-2.05)
  {Shared error\\FIFO ($e$)};
\draw[flow] (6.45,0 |- entry.south)--(6.45,-2.05)--(errorfifo.west);
\node[finite,text width=2.25cm,minimum height=0.7cm] (ownederror) at (11.05,-2.05)
  {Paired error\\group};
\draw[flow] (errorfifo.east)--(ownederror.west);
\draw[flow] (ownederror.north)--(convert.south);
\node[mem,text width=1.6cm,minimum height=1.4cm] (hostout) at (1.0,-3.48)
  {Host output\\Six $c_0$ limbs\\Assemble\\$(c_0,c_1)$};
\node[finite,text width=1.1cm] (outmem) at (4.2,-3.48) {DMA\\writer};
\node[finite,text width=1.55cm] (packetizer) at (7.65,-3.48)
  {Common $c_0$\\packetizer};
\node[finite,text width=4.7cm,minimum height=1.1cm] (moduli) at (11.25,-3.48)
  {Six modulus pipelines $q_0,\ldots,q_5$\\Each: 2 NTTs, arithmetic, $c_0$ FIFO\\One pipeline expanded in (b)};
\draw[flow] (fanout.south)--(14.75,-3.48)--(moduli.east);
\draw[flow] (5.8,0 |- entry.south)--(5.8,-2.65)--
  node[above,labeltext,pos=0.18]{$s_i$, $c_{1,i}$}(11.25,-2.65)--(moduli.north);
\draw[flow] (moduli.west)--(packetizer.east);
\draw[flow] (packetizer.west)--(outmem.east);
\draw[flow] (outmem.west)--(platform.east |- outmem.west);
\draw[flow] (platform.west |- hostout.east)--(hostout.east);
\draw[flow,dashed] (hostin.south)--node[right,labeltext]{Retained\\$c_1$}(hostout.north);
\node[anchor=west,font=\footnotesize] at (0,-4.52)
  {Host control through OPAE SDK. One complete operation at a time; four coefficients per internal transfer.};
\node[anchor=west,font=\small\bfseries] at (0,-5.1)
  {(b) One physical modulus pipeline $q_i$ (six spatial copies)};
\draw (3.0,-5.68) rectangle (8.2,-6.98);
\node[anchor=west,labeltext] at (3.15,-5.68) {Message NTT task};
\node[finite,text width=1.85cm] (reduce) at (4.15,-6.35)
  {Signed Barrett\\modulo $q_i$};
\node[rtl,text width=1.75cm] (nttb) at (6.95,-6.35)
  {NTT-B\\Message + error};
\node[rtl,text width=1.75cm] (ntta) at (6.95,-7.62)
  {NTT-A\\Secret key};
\node[finite,text width=3.0cm,minimum height=1.35cm] (poly) at (11.1,-6.98)
  {Ciphertext arithmetic\\$c_{0,i}=\widehat{m+e}_i$\\${}-\widehat{s}_i\odot c_{1,i}\pmod{q_i}$};
\node[mem,text width=1.7cm] (exit) at (14.45,-6.98) {$c_{0,i}$ FIFO};
\node[anchor=west,font=\footnotesize,align=left] (shared) at (0,-6.35)
  {Signed message\\from shared fanout};
\draw[flow] (shared.east)--(reduce.west);
\draw[flow] (reduce.east)--(nttb.west);
\node[anchor=west,font=\footnotesize] (secret) at (0,-7.62) {$s_i$ residues from input};
\draw[flow] (secret.east)--(ntta.west);
\draw[flow] (nttb.east)--(8.8,-6.35)--(8.8,-6.6)--(poly.west |- {0,-6.6});
\draw[flow] (ntta.east)--(8.8,-7.62)--(8.8,-7.36)--(poly.west |- {0,-7.36});
\node[anchor=west,font=\footnotesize] (uniform) at (0,-8.5) {NTT-form $c_{1,i}$ from input};
\draw[flow] (uniform.east)--(11.1,-8.5)--(poly.south);
\draw[flow] (poly.east)--(exit.west);
\draw[flow] (exit.east)--(16.05,-6.98)--(16.05,-7.55);
\node[anchor=north,font=\footnotesize,align=center] at (16.05,-7.62)
  {To common\\packetizer};
\node[anchor=west,font=\footnotesize,align=left,text width=16.3cm] at (0,-9.22)
  {The six modulus pipelines produce the limbs of one ciphertext. Application clock: 470.8 MHz.\\
   Thick borders: DSP Builder transform RTL. Other AFU blocks: SYCL tasks, storage, and DMA logic.};
\end{tikzpicture}
\caption{CKKS client accelerator dataflow. (a) OFS FIM/PIM connects host memory to the AFU's DMA engines. The SYCL pipeline generated with Altera HLS IP Gen compiler shares the IFFT, conversion, and error addition across six modulus paths. Their $c_0$ limbs return through one packetizer for assembly with retained $c_1$. (b) Each modulus has independent secret-key and message NTTs; signed Barrett reduction precedes the message transform. Four coefficient lanes carry parts of one polynomial.}
\label{fig:full-accelerator-overview}
\end{figure*}
One operation consumes a message vector and produces one complete ciphertext. The AFU admits one operation at a time while its internal arithmetic stages process coefficient streams concurrently. Within an operation, each transform accepts four coefficients per transfer: these are the four \emph{coefficient lanes}. The six \emph{modulus pipelines}, by contrast, process six representations of the same polynomial, one for each prime.

In Fig.~\ref{fig:full-accelerator-overview}(a), a DMA reader streams prepared inputs from host memory into the input parser. The encoding vector traverses the shared IFFT, followed by scaling, rounding, signed conversion, and error addition. A FIFO stores the converted coefficients for distribution to the six modulus pipelines. Secret-key coefficients and NTT-form \(c_1\) bypass the IFFT. Each modulus pipeline reduces the shared coefficients, computes the two NTTs, and forms its \(c_0\) limb as expanded in Fig.~\ref{fig:full-accelerator-overview}(b). Dedicated engines let the modulus computations proceed without sharing a transform scheduler. A common output packetizer collects the six \(c_0\) limbs for DMA delivery to host memory. The host combines them with retained \(c_1\).

\subsection{SYCL and the Hardware/Software Partition}

DSP Builder generates the computationally intensive IFFT and NTT datapaths as RTL cores \cite{DSPBuilder}. We package these cores as imported libraries and call them from SYCL tasks. Other tasks parse and distribute input, convert and broadcast coefficients, apply modular reduction, compute ciphertext arithmetic, and collect output. Altera HLS IP Gen compiler generates the hardware pipeline from this SYCL implementation, with FIFO pipes carrying intermediate coefficient groups on chip \cite{HLSIPGen}.

The generated pipeline IP is integrated into an AFU within Open FPGA Stack (OFS)~\cite{OFSOverview}. The FPGA Interface Manager (FIM) supplies the platform interfaces, clocks, and resets~\cite{OFSProductBrief}; the Platform Interface Manager (PIM) adapts its host channel to the AFU's memory and control interfaces~\cite{OFSPIM}. The AFU contains the DMA engines, controller, and generated SYCL pipeline. A C++ host program prepares inputs with SEAL-Embedded and uses the OPAE SDK, the software development kit for the Open Programmable Acceleration Engine framework, to access the AFU and register host-memory buffers~\cite{OPAEGuide}. The AFU's RTL handles DMA and control; the generated pipeline combines DSP Builder transform RTL with SYCL-defined arithmetic and dataflow. The hardware pipeline remains resident on the FPGA. The AFU invokes it once per encryption operation, with its internal tasks executing concurrently.

\subsection{Input Buffering and Shared IFFT}

The input stage separates host-memory transfer from delivery to the arithmetic pipelines. The DMA reader supplies 512-bit words, which the parser assembles into groups containing four encoding values, their error samples, and the secret-key and \(c_1\) coefficients for every modulus. A FIFO between the parser and distributor can hold a complete polynomial's input groups. The distributor sends each field to its consumer in coefficient order. Blocking pipe writes hold the current group until its destinations accept it, accommodating differences in consumer progress without overwriting unread input.

The imported IFFT RTL IP core is based on a Gentleman--Sande decimation-in-frequency butterfly \cite{gentleman1966fast}:
\begin{align}
	u &= a+b
	\label{eq:ifft_u}, \\
	v &= (a-b) \cdot \omega,
	\label{eq:ifft_v}
\end{align}
where \(a\), \(b\), and \(\omega\) are complex doubles. The four coefficient lanes carry one group of four complex values per accepted transfer, so a length-\(N\) vector requires 2,048 transfers.

The IFFT task pairs each input group with its error samples. As transform results return, it applies \(\Delta/N\), rounds the real components, converts them to signed 64-bit integers, and adds the errors. A FIFO sized for at least one complete polynomial holds these four-coefficient, 256-bit groups. The broadcast task reads each group once and delivers a copy to every modulus pipeline before advancing. Thus, the floating-point transform and integer conversion are shared across all six moduli, while the FIFO accommodates short-term differences in consumer progress.

\subsection{DSP Builder Streaming Transform Architecture}
\label{DSP Builder Streaming Transform Architecture}

Fig.~\ref{fig:dsp-builder-pipelines} compares the IFFT and NTT pipelines and expands the transpose-memory mechanism that aligns operands between butterfly sections.
\begin{figure*}[t]
\centering
\begin{tikzpicture}[x=1cm,y=1cm,>=stealth,
  block/.style={draw,minimum height=0.85cm,align=center,font=\footnotesize,inner sep=3pt},
  route/.style={block,fill=black!6},
  arithmetic/.style={block,fill=black!5},
  note/.style={font=\footnotesize,align=center},
  flow/.style={->,thick}]
\node[anchor=west,font=\small\bfseries] at (0,1.05)
  {(a) Negacyclic IFFT: four complex doubles per transfer};
\node[route,text width=1.65cm] (ip) at (0.9,0) {Input transpose\\+ conjugate};
\node[arithmetic,text width=1.3cm] (i2) at (2.9,0) {Two parallel\\radix-2 BFs};
\node[arithmetic,text width=1.4cm] (it) at (5.05,0) {General\\twiddle};
\node[route,text width=1.2cm] (ix) at (6.95,0) {Transpose};
\node[arithmetic,text width=1.3cm] (i4) at (8.85,0) {Radix-4\\butterfly};
\draw[dashed] (4.05,-0.68) rectangle (9.8,0.65);
\node[note,fill=white,inner sep=1pt] at (6.925,-0.68) {Six spatial sections in series};
\node[route,text width=1.3cm] (io) at (11.1,0) {Final\\transpose};
\node[arithmetic,text width=1.6cm] (ic) at (13.3,0) {Conjugate +\\negacyclic phase};
\node[note,text width=2.1cm] (is) at (15.75,0)
  {SYCL adapter\\Scale, round\\Add error $e$};
\foreach \a/\b in {ip/i2,i2/it,it/ix,ix/i4,i4/io,io/ic,ic/is}
  \draw[flow] (\a.east)--(\b.west);
\node[anchor=west,font=\small\bfseries] at (0,-1.5)
  {(b) Negacyclic NTT: four coefficients per transfer, with the modulus selector aligned with data};
\node[route,text width=1.65cm] (np) at (0.9,-2.5) {Input\\transpose};
\node[arithmetic,text width=1.3cm] (nw) at (2.9,-2.5) {NWC twiddle\\+ reduction};
\node[arithmetic,text width=1.4cm] (n4) at (5.05,-2.5) {Radix-4\\modular BF};
\node[arithmetic,text width=1.2cm] (nt) at (6.95,-2.5) {Twiddle +\\reduction};
\node[route,text width=1.3cm] (nx) at (8.85,-2.5) {Transpose};
\draw[dashed] (4.05,-3.2) rectangle (9.8,-1.85);
\node[note,fill=white,inner sep=1pt] at (6.925,-3.2) {Six spatial sections in series};
\node[arithmetic,text width=1.3cm] (n2) at (11.1,-2.5) {Two parallel\\radix-2 BFs};
\node[arithmetic,text width=1.6cm] (nr) at (13.3,-2.5) {Final coarse /\\fine reduction};
\node[note,text width=2.1cm] (no) at (15.75,-2.5)
  {30-bit residues\\32-bit words\\to SYCL pipe};
\foreach \a/\b in {np/nw,nw/n4,n4/nt,nt/nx,nx/n2,n2/nr,nr/no}
  \draw[flow] (\a.east)--(\b.west);
\node[anchor=west,font=\small\bfseries] at (0,-3.95)
  {(c) Transpose structure: permutation across lanes and transfer times};
\node[route,text width=2.0cm] (wm) at (2.1,-4.8) {Write-side\\lane routing};
\node[route,text width=3.2cm] (ram) at (6.7,-4.8) {Banked dual-port memories\\Independent write/read addresses};
\node[route,text width=2.0cm] (rm) at (11.3,-4.8) {Read-side\\lane routing};
\draw[flow] (0,-4.8)--node[above,font=\scriptsize]{4 lanes}(wm.west);
\draw[flow] (wm.east)--(ram.west);
\draw[flow] (ram.east)--(rm.west);
\draw[flow] (rm.east)--node[above,font=\scriptsize]{4 lanes}(13.5,-4.8);
\node[note,anchor=west,text width=2.6cm,font=\scriptsize] at (13.7,-4.8)
  {Memory depth varies by stage.\\NTT words also carry the modulus selector.};
\end{tikzpicture}
\caption{DSP Builder IFFT and NTT pipelines and their transpose memories. BF denotes butterfly and NWC denotes negacyclic wraparound convolution. Each dashed group represents six spatial sections in series. The IFFT places radix-2 first and applies the negacyclic phase after the Fourier pipeline. Its SYCL output conversion scales by $\Delta/N$, rounds, and adds $e$ before broadcasting signed coefficients. The NTT places its NWC layer before the butterflies and radix-2 last. Panel (c) routes coefficients across lanes and transfer times, with memory widths and depths set by the stage. Import buffering is described in Section~\ref{sec:rtl-integration}.}
\label{fig:dsp-builder-pipelines}
\end{figure*}

DSP Builder generates pipelines of butterfly sections, twiddle-factor arithmetic, and memory-based permutations. The 13 radix-2 stages are grouped into seven mixed-radix sections using \(8192=2\cdot4^6\). Interstage routing supplies the operands needed by each section while keeping intermediate polynomials in on-chip storage.

\subsubsection{Negacyclic IFFT}
The IFFT begins with an input transpose and complex conjugation. Two parallel radix-2 butterflies feed six four-point butterfly sections, connected by general-twiddle blocks and transpose networks. A final transpose reorders the result. Within a four-point section, complex sums, differences, and fixed rotations implement the butterfly. The coefficient tables supply the general-twiddle multiplications between sections. All real and imaginary components use double-precision arithmetic. Conjugation at the output completes the unnormalized inverse transform, followed by a lane-specific, ROM-driven phase multiplication for the negacyclic transform. The SYCL stage then applies \(\Delta/N\).

\subsubsection{Negacyclic NTT}
The NTT begins with a transpose and a negacyclic twiddle multiplication/reduction layer. Six radix-4 butterfly sections alternate with general-twiddle and transpose blocks, followed by two parallel radix-2 butterflies. A modulus selector travels with the data and selects the twiddle factors and reduction constants for \(q_i\). The butterfly adders and subtractors form integer sums and differences of their input coefficients. They retain extra bits for intermediate results, reduction around the twiddle multiplications limits their range, and final coarse/fine correction produces residues in \([0,q_i)\). The generated structure supports all six configured primes and is instantiated for each transform role in each modulus pipeline.

The transpose networks supply coefficients that arrive on different lanes or at different times to the same butterfly section. Lane-selection multiplexers reorder values within a transfer. Independently addressed dual-port memories hold earlier coefficients and read them with later arrivals in the order required by the next section. This storage implements permutations that cannot be completed by exchanging the four values of one transfer alone. Valid-data control and the NTT modulus selector remain aligned with the reordered coefficients.

\subsection{Parallel Modulus Pipelines}

Let \(j\) index a coefficient, \(x_j\) denote an IFFT output, and \(e_j\) denote the shared error sample. The shared IFFT task computes \(t_j\) once. Each modulus pipeline reduces it to the plaintext-plus-error coefficient \(p_{i,j}\):
\begin{align}
 t_j &=\operatorname{round}\!\left(\operatorname{Re}(x_j) \cdot \frac{\Delta}{N}\right)+e_j,
 \label{eq:sr_t} \\
 p_{i,j} &= t_j \pmod{q_i}.
 \label{eq:sr_p}
\end{align}
Signed Barrett reduction in the message-input adapter reduces \(t_j\) modulo \(q_i\). NTT-B in Fig.~\ref{fig:full-accelerator-overview}(b) transforms the resulting plaintext-plus-error stream, while NTT-A independently transforms the secret-key stream. Their outputs and the supplied \(c_{1,i}\) are buffered until corresponding coefficient groups can be combined.

Using the NTT notation of \eqref{eq:c_0}, the ciphertext arithmetic kernel computes
\begin{equation}
 \label{eq:comp_c0}
 c_{0,i,j}=\widehat{m+e}_{i,j}-\widehat{s}_{i,j}\,c_{1,i,j}\pmod{q_i}.
\end{equation}
Thus \(c_{0,i}\) in Fig.~\ref{fig:full-accelerator-overview} denotes the whole limb for modulus \(i\), and \(c_{0,i,j}\) in \eqref{eq:comp_c0} denotes its coefficient \(j\). All six modulus pipelines have separate NTT-A and NTT-B engines, giving twelve NTT engines in total. Complete-vector buffers align their outputs. Each pipeline writes its \(c_0\) groups into a dedicated FIFO, and a common packetizer drains the limbs in modulus order.

\subsection{RTL Integration and Flow Control}
\label{sec:rtl-integration}

The imported transforms produce output that their downstream consumers may not immediately accept. A wrapper between each RTL core and its SYCL kernel packs coefficient groups, marks valid data, and buffers results until the next stage is ready. The C++ declaration and RTL import specification agree on the placement of the four coefficients and their control fields.

\subsubsection{SYCL Pipeline Operation}
The AFU starts one generated hardware pipeline for each admitted operation. Within that component, SYCL tasks implement input distribution, transforms, broadcast, modular reduction, ciphertext arithmetic, and output collection. The tasks communicate through blocking pipes and execute concurrently as data and output space become available. Their operation is coordinated in hardware; the host waits for the AFU's complete-operation result.

The IFFT and NTT tasks each wait for an input coefficient group before calling the imported RTL. Their wrappers capture valid transform outputs and deliver them as downstream consumers accept them. In modulus pipeline \(i\), both NTT engines use the fixed selector for \(q_i\), with a distinct prime assigned to each pipeline.

\subsubsection{Burst Capture and Output-Space Reservation}
Each import wrapper has a 4,096-entry block-memory output FIFO, enough for two transforms of 2,048 coefficient groups each. This FIFO is separate from the complete-vector buffers between SYCL kernels. A nonempty FIFO presents its first entry until the downstream interface accepts it. Counters track accepted input groups and fully drained outputs to reserve space for results still in flight.

When the next stage stalls, the wrapper holds buffered output until it is accepted. It does not rely on directly stopping the generated arithmetic. The compiler treats each imported component as a variable-latency interface. Transfers follow valid/ready signals rather than a fixed cycle count. The downstream SYCL task forwards each completed return into its output pipe. This separates the transform's output burst from the rate at which later arithmetic can consume it, while preserving the four-coefficient grouping and order.

\subsection{Host Runtime and Ciphertext Assembly}

The host allocates and registers reusable input and output regions in host memory. For each operation, it copies the prepared input into the registered region, supplies the AFU with the DMA addresses and payload lengths, and starts execution through memory-mapped control registers. The DMA reader feeds the SYCL pipeline directly, and the DMA writer returns the packetized \(c_0\) limbs to the registered output region. Intermediate transform data remains in on-chip buffers.

Completion requires both the arithmetic pipeline and output transfer to finish. The AFU applies a write-ordering fence before reporting the output ready, so the host can read the returned coefficients before reusing the registered regions. Collection expands the 32-bit residues into the ciphertext storage layout used by Microsoft SEAL and combines all six returned \(c_0\) limbs with the retained \(c_1\) limbs. The resulting ciphertext contains both polynomials in NTT representation with the configured scale and modulus chain. The measured host path repeats this sequence one complete ciphertext at a time.

\section{Experimental Results}

\subsection{Experimental Setup}
\label{sec:experimental-setup}

The accelerator runs at 471 MHz on a BittWare IA-840F card with an Altera Agilex 7 AGFB027R25A2E2V FPGA. The software reference is one native SEAL-Embedded worker on the same Intel Xeon W-2295 workstation. The implementation uses the Altera HLS IP Gen compiler 2026.1, Quartus Prime Pro 26.1.1, and OFS FIM 2026.1-1. Host access uses OPAE SDK 2.14.0-3.

Both implementations use \(N=8192\), scale \(2^{25}\), the same key pair, and the six primes in Table~\ref{tab:rns-moduli}, taken from the SEAL-Embedded source code \cite{SEALEmbeddedGit}. The accelerator uses 16 prepared real, partial, and complex cases; the CPU uses 16 real-valued variants with all 4,096 slots populated. Both receive prepared embeddings, sampled errors, uniform polynomials, and coefficient-domain keys. The input sets and delivery paths differ, so the CPU measurements provide a software reference rather than an identical-input speedup comparison.

\begin{table}[t]
\centering\footnotesize
\caption{Fixed Primes for the Six Modulus Pipelines at \(N=8192\), from the SEAL-Embedded Source Code \cite{SEALEmbeddedGit}}
\label{tab:rns-moduli}
\begin{tabular}{cr}
\toprule
Pipeline \(i\) & Prime \(q_i\) \\
\midrule
0 & 1,053,818,881 \\
1 & 1,054,015,489 \\
2 & 1,054,212,097 \\
3 & 1,055,260,673 \\
4 & 1,056,178,177 \\
5 & 1,056,440,321 \\
\bottomrule
\end{tabular}
\end{table}

The host timer covers IFFT, scaling and rounding, error addition, RNS reduction, both NTTs, ciphertext arithmetic, and complete ciphertext construction. FPGA timing also includes prepared-input copying, submission, completion, output delivery, and per-operation deadline checks. Input preparation, allocation, warmup, and numerical verification are outside timing. The CPU copies \(c_1\) into its output storage; the FPGA host assembles the returned \(c_0\) with retained \(c_1\).

Output storage is allocated before timing. The drivers retain the first, middle, and last outputs of each measured trial by exchanging buffer ownership. Ciphertext construction remains timed; decryption and decoding with unmodified Microsoft SEAL follow the timed interval.

The accelerator runs three trials at each of 1,024 and 8,192 operations. Each process qualifies the prepared corpus and completes 32 warmup operations before timing. The CPU runs three fresh processes at each of 1,024 and 8,192 operations, with 64 warmup operations per process. Reported rates divide total complete ciphertexts by total measured seconds.

\subsection{Resource Utilization and Correctness}

Table~\ref{tab:post-fit} reports the fitted resource attribution for the measured accelerator. The resources include the computational pipeline, DMA, control, and packet adaptation contained within the AFU region along with the platform infrastructure provided by the OFS FIM/PIM. Block-memory bits and M20K counts describe storage capacity and physical blocks, respectively.

\begin{table}[t]
\centering\footnotesize
\caption{Resource Attribution for the Accelerator and OFS Platform}
\label{tab:post-fit}
\begin{tabular}{@{}lrrr@{}}
\toprule
Metric & Total \\
\midrule
ALMs needed & 476,229 \\
Logic ALUTs & 538,822 \\
Registers & 1,071,132 \\
Block-memory kbits & 81,357 \\
M20Ks & 4,442 \\
DSPs needed & 1,856 \\
\bottomrule
\end{tabular}
\par\smallskip
\end{table}

The accelerator passes all numerical qualification tests. Every ciphertext passes all 4,096 decoded slots, including inactive slots, with canonical \(c_0/c_1\) residues and exact retained-\(c_1\) identity. Maximum absolute decoded error is \(3.07\times10^{-5}\) for the accelerator and \(3.18\times10^{-5}\) for the CPU, below the 0.1 threshold.

Numerical coverage excludes warmup outputs and unsampled measured outputs. The \(c_0\) checks assess canonicality and use unmodified Microsoft SEAL for decryption and decoding.

\subsection{Throughput}
\label{sec:matched-throughput}

\emph{Throughput} is the number of complete ciphertexts returned per second over a long sequence, including the host work required to submit computation and construct outputs. One complete ciphertext contains both components across all six moduli.

Each record denotes one encoding-and-encryption operation. Table~\ref{tab:matched-performance} compares the runs of 1,024 and 8,192 records. FPGA / CPU is the ratio of the unrounded weighted rates, giving 45.25\(\times\) and 45.35\(\times\), respectively, for the stated workloads.

\begin{table*}[t]
\centering\footnotesize
\caption{Throughput Comparison of CPU and FPGA Implementations}
\label{tab:matched-performance}
\setlength{\tabcolsep}{2pt}
\begin{tabular}{>{\raggedright\arraybackslash}p{0.155\textwidth}>{\raggedright\arraybackslash}p{0.155\textwidth}>{\raggedright\arraybackslash}p{0.155\textwidth}>{\raggedright\arraybackslash}p{0.155\textwidth}>{\raggedright\arraybackslash}p{0.155\textwidth}>{\raggedright\arraybackslash}p{0.155\textwidth}}
\toprule
Records per run & CPU weighted ct/s & CPU trial range & FPGA weighted ct/s & FPGA trial range & FPGA / CPU \\
\midrule
1,024 & 113.42 & 113.38--113.46 & 5,133 & 5,125--5,139 & 45.25\(\times\) \\
8,192 & 113.43 & 113.38--113.48 & 5,144 & 5,057--5,195 & 45.35\(\times\) \\
\bottomrule
\end{tabular}
\end{table*}

\subsection{Comparison with Prior Custom Accelerators}
\label{sec:prior-comparison}

Table~\ref{tab:prior-accelerators} compares the operation scope and reported results of client custom accelerators. The following discussion relates their arithmetic organization to our dedicated encoding/encryption path.

\begin{table*}[t]
\centering\footnotesize
\caption{Client Accelerator Architectures and Measurement Boundaries}
\label{tab:prior-accelerators}
\setlength{\tabcolsep}{2pt}
\begin{tabular}{>{\raggedright\arraybackslash}p{0.112\textwidth}>{\raggedright\arraybackslash}p{0.27\textwidth}>{\raggedright\arraybackslash}p{0.14\textwidth}>{\raggedright\arraybackslash}p{0.158\textwidth}>{\raggedright\arraybackslash}p{0.251\textwidth}}
\toprule
System & Operation and parameters & Platform & Reported result & Measurement scope \\
\midrule
This work & Dedicated IFFT/NTTs; symmetric encoding/encryption; \(N=8192\); six concurrent 30-bit moduli; prepared real/complex inputs & Altera Agilex 7; 471 MHz & 5,143 complete ciphertexts/s & Prepared-input copy through AFU execution, output delivery, and host ciphertext construction; input preparation excluded \\
Aloha-HE \cite{draft4} & Public-key encoding/encryption; \(N=8192\), \(3\times54\)-bit & AMD Kintex-7 / MicroBlaze; 200 MHz & 1.87 ms & Whole-system latency: software, hardware, and transfers \\
Lee \emph{et al.}~\cite{draft11} & Public-key encryption; \(N=8192\), level 5; widths \{30,20,20,20,20,20,30\} bits & AMD XCU250; 250 MHz & 2,950 encryption operations/s & Encryption-module rate, starting with an encoded polynomial \\
ABC-FHE \cite{draft15} & Encoding/encryption; \(N=65536\), 36-bit primes, 24 levels & 28-nm ASIC design; 600 MHz & 1,112\(\times\) over Core i7-12700 / Lattigo & Cycle model; absolute latency shown only in a plot \\
\bottomrule
\end{tabular}
\end{table*}

Aloha-HE shares one FFT butterfly and three NTT butterflies across encoding, encryption, decryption, and decoding to reduce area \cite{draft4}. Our one double-precision IFFT and twelve NTT engines keep the encoding/encryption transforms physically separate. This removes transform scheduling conflicts between the two NTT roles and the six moduli, at the cost of more resident arithmetic. Lee \emph{et al.}'s configurable encryption/decryption hardware accepts an already encoded polynomial for the reported encryption rate \cite{draft11}. Our pipeline also performs the preceding encoding and passes its coefficients directly to encryption on chip.

ABC-FHE already exploits parallel streaming. Its two reconfigurable cores share arithmetic between FFT and NTT modes, use custom 55-bit floating point, and generate randomness and twiddles on chip \cite{draft15}. Our fixed symmetric configuration retains double-precision encoding output and dedicates each NTT to one role and modulus, with host-prepared randomness and ROM twiddles. SYCL connects those independent RTL paths to the surrounding arithmetic and DMA streams. ABC-FHE supports larger bootstrappable parameters and evaluates them with an ASIC cycle model. Our results measure a physical FPGA at the stated fixed configuration.

In both our design and Aloha-HE, the modulus limbs collectively represent one ciphertext. Our host-observed throughput, Aloha-HE's whole-system latency, Lee \emph{et al.}'s encryption-module rate, and ABC-FHE's cycle model use different operation scopes and parameters. Table~\ref{tab:prior-accelerators} therefore provides architectural context rather than a normalized performance ranking.

\subsection{Discussion and Limitations}

The accelerator sustains approximately 5,143 complete ciphertexts/s while admitting one operation at a time. Within each operation, the shared IFFT feeds dedicated, concurrent NTT paths for all six moduli. The measurements include host submission and output construction, so they characterize the complete accelerator path rather than an isolated transform. Stage-level measurements are needed to determine which arithmetic or transfer stage limits this rate.

Larger polynomial degrees require regenerated transforms, coefficient tables, permutations, and storage. Longer modulus chains could reuse the IFFT result across successive modulus groups, with additional replay storage and aligned auxiliary streams. Wider primes require revised arithmetic widths, reduction logic, and interfaces. These extensions remain implementation and validation work.

\section{Conclusion}

The accelerator connects one shared IFFT with four coefficient lanes and twelve dedicated NTT engines across six modulus pipelines through SYCL pipes. DSP Builder supplies the transform arithmetic, and Altera HLS IP Gen compiler generates the surrounding pipeline for an OFS-integrated AFU. At 471 MHz on Agilex 7, the accelerator delivers 5,143 complete ciphertexts/s with input preparation excluded. One SEAL-Embedded CPU worker on the same workstation delivers 113 ciphertexts/s for real-valued inputs using all 4,096 slots. All qualification ciphertexts and retained measured samples pass complete-ciphertext and all-slot checks. Future work will extend the architecture to larger polynomial degrees and wider RNS primes, requiring regenerated transforms, revised modular arithmetic, and evaluation at the new parameters.

\section*{Acknowledgments}
%
This material is based upon work supported by the National Science Foundation under Grant No. 2347253.
The authors would like to thank Altera Corporation for providing the FPGA development board used in this work.


\end{document}